\documentclass[useAMS,usenatbib]{mn2e}
\usepackage{graphicx}
\usepackage{amssymb}
\usepackage{enumerate}
\usepackage{times}

\footnotesize
\newdimen\digitwidth    
\setbox0=\hbox{\rm0}
\digitwidth=\wd0
\catcode`!=\active
\def!{\kern\digitwidth}
\normalsize

\newcommand{\psr}{PSR~J2051$-$0827}

\title[Quadrupole variations in the J2051$-$0827 system]{Evidence for
  gravitational quadrupole moment variations in the companion of \psr}

\author[K.Lazaridis et al.]{K.~Lazaridis,$^1$ J.~P.~W.~Verbiest,$^1$
  T.~M.~Tauris,$^2$$^,$$^1$ B.~W.~Stappers,$^3$$^,$$^4$
  M.~Kramer,$^1$$^,$$^3$\newauthor \ N.~Wex,$^1$ A.~Jessner,$^1$
  I.~Cognard,$^{5,6}$ G.~Desvignes,$^5$ G.~H.~Janssen,$^3$
  M.~B.~Purver,$^{3}$ \newauthor \ G.~Theureau,$^{5,6}$ C.~G.~Bassa,$^3$
  R.~Smits$^{3,4}$
  \\
  $^{1}$Max-Planck-Institut f\"ur Radioastronomie, Auf dem H\"ugel 69,
  53121 Bonn, Germany
  \\
  $^{2}$Argelander-Institut f\"ur Astronomie, Universit\"at Bonn, Auf
  dem H\"ugel 71, 53121 Bonn, Germany 
  \\
  $^{3}$University of Manchester, Jodrell Bank Centre for
  Astrophysics, Alan Turing Building, Manchester M13 9PL, UK
  \\
  $^{4}$Stichting ASTRON, Postbus 2, 7990 AA Dwingeloo, The
  Netherlands
  \\
  $^{5}$Laboratoire de Physique et Chimie de l'Environnement, CNRS, 3A
  Avenue de la Recherche Scientifique, \\ 
  $\,$ 45071 Orl\'{e}ans Cedex 2, France
  \\
  $^{6}$Station de Radioastronomie de Nan\c cay, Paris Observatory, France}
\date{Drafted 22 November 2010.}

\begin{document}

\maketitle

\begin{abstract} 
  We have conducted radio timing observations of the eclipsing
  millisecond binary pulsar J2051$-$0827 with the European Pulsar
  Timing Array network of telescopes and the Parkes radio telescope,
  spanning over 13\,years. The increased data span allows significant
  measurements of the orbital eccentricity, $e=(6.2 \pm 1.3)\times
  10^{-5}$ and composite proper motion, $\mu_t=7.3 \pm
  0.4$\,mas\,yr$^{-1}$. Our timing observations have revealed secular
  variations of the projected semi-major axis of the pulsar orbit
  which are much more extreme than those previously published; and of
  the orbital period of the system. Investigation of the physical
  mechanisms producing such variations confirm that the variations of
  the semi-major axis are most probably caused by classical spin-orbit
  coupling in the binary system, while the variations in orbital
  period are most likely caused by tidal dissipation leading to
  changes in the gravitational quadrupole moment of the companion.
\end{abstract}


\section{Introduction}
\label{sec:Intro}

\psr\ is the second eclipsing millisecond pulsar system discovered
after the original black widow pulsar PSR~B1957$+$20 \citep{fst88}. It
was discovered \citep{sbl+96} as part of a Parkes all-sky survey of
the southern sky for low-luminosity and millisecond pulsars. The
pulsar has a spin period of 4.5\,ms and inhabits a very compact
circular orbit with a very low mass companion: $m_{\rm c} <
0.1\,$M$_{\odot}$.  The orbital period is $P_{\rm b} = 2.38$\,h and
the pulsar and its companion are separated by just
1.0\,R$_{\odot}$. \cite{sbl+96, sbl+01a} investigated the eclipse
characteristics of this system and found that their duration is $\sim
10$\% of the orbital period at frequencies below 1\,GHz. In addition,
assuming a mean free electron density of $10^7$\,cm$^{-3}$ in the
eclipse region, they calculated a mass-loss rate of $\sim
10^{-14}$\,M$_\odot$\,yr$^{-1}$, which is insufficient to evaporate
the companion completely within a Hubble time.

\begin{figure*}
\begin{center}
  \mbox{\includegraphics[width=0.70\textwidth, angle=0]{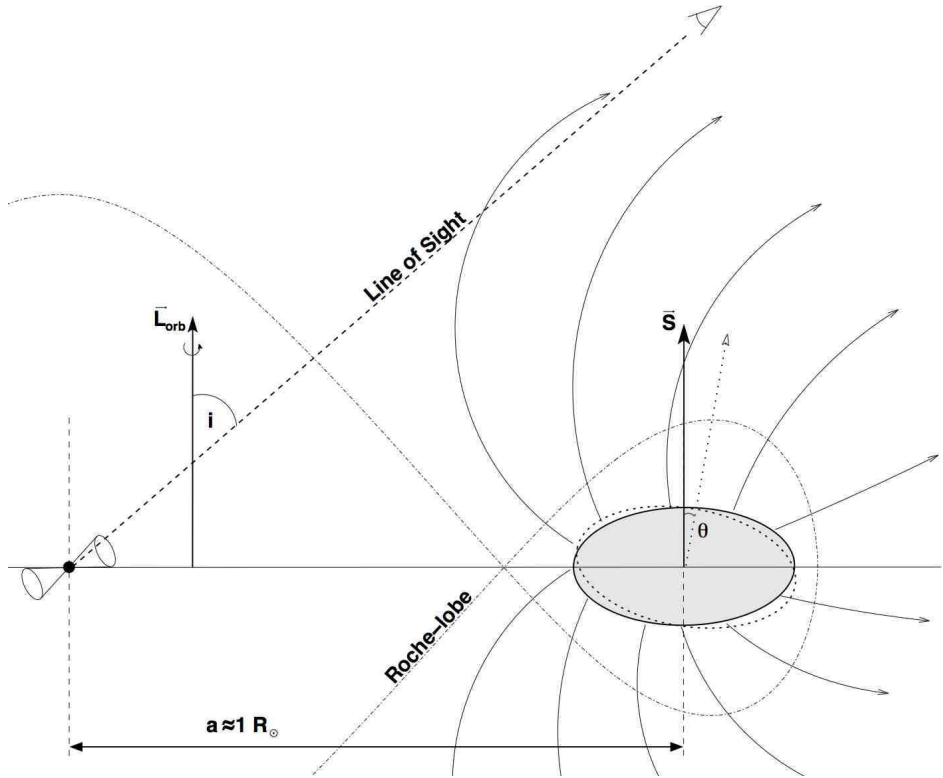}}
  \caption{Illustration of the J2051-0827 system geometry. The pulsar
    and companion star are about one solar radius apart, the
    inclination angle of their orbit is approximately $40^{\circ}$ and
    the pulsar wind pushes the companion's wind out, resulting in a
    bow shock-like denser region of charged particles, which causes
    the system to be eclipsed at given orbital phases and low
    observational frequencies. Changes in the oblateness of the
    tidally-locked companion star, possibly in combination with
    alterations in the direction of the spin axis $\theta$, trigger
    changes in the quadrupole moment that affect the timing of
    the pulsar. Notice this sketch is for illustrative purposes only
    and is not to scale.}
\label{fig:Geom}
\end{center}
\end{figure*}

Optical observations of the companion revealed that it is irradiated
by the pulsar wind \citep{sbb96}. Further observations and modelling
of the variability of the light curve \citep{svlk99,svbk01b}
determined a binary inclination angle of $i \sim 40^{\circ}$, a
backside temperature of $T \leq 3000$\,K, a companion mass of $\sim
0.04$\,M$_\odot$ and a radius $\sim 0.064$\,R$_\odot$. An alternative
model with a companion almost filling its Roche lobe ($\sim
0.12$\,R$_\odot$) was also considered and produced almost the same
results, albeit with a significantly worse fit. For clarification
purposes, a cartoon of the system is given in Figure~\ref{fig:Geom}.

\cite{dlk+01} presented the most precise timing analysis of \psr\
using $\sim 6$\,yr of radio timing measurements with the Effelsberg
100-m radio telescope and the 76-m Lovell at Jodrell Bank. Among the
most important measurements is the variation of the projected
semi-major axis, probably caused by Newtonian spin-orbit coupling
in this binary system. In addition, they measured significant orbital
period variations and concluded that those were created by the same
mechanism as in the B1957$+$20 system \citep{as94, aft94}: by
tidal dissipation of a tidally powered, non-degenerate companion.

Because black widow (and other eclipsing) systems are susceptible to
variations of their orbital parameters, monitoring over long lengths
of time is crucial to understand these systems. Timing of black widow
pulsars has been undertaken by \citet{nat00} and
\citet{fck+03,fhn+05}, but only very few of these systems have been
monitored for a decade or longer. This is of particular interest given
the large number of black widow pulsars recently found in unidentified
\emph{Fermi} sources \citep[and several discoveries that are soon to
  be published]{rrc+11,kjr+11}.

In this paper we revisit \psr\ with 13 years of high-precision timing
data and combined datasets from the Australia Telescope National
Facility's Parkes 64-m radio telescope and the European Pulsar Timing
Array (EPTA) telescopes, consisting of the 100-m Effelsberg
radio-telescope of the Max-Planck-Institute for Radioastronomy,
Germany, the 76-m Lovell radio-telescope at Jodrell Bank observatory
of the University of Manchester, UK, the 94-m-equivalent Westerbork
Synthesis Radio Telescope (WSRT) of ASTRON, the Netherlands, and the
94-m-equivalent Nan\c cay decimetric Radio Telescope (NRT) of the
CNRS, France. After describing the properties of our multi-telescope
data (Section~\ref{sec:obs}), we present the updated measurements of
the astrometric, spin and binary parameters for the system
(Section~\ref{sec:timing}). Specifically we show the measurement of
the combined proper motion and of the orbital eccentricity of the
\psr\ system for the first time and we measure the dispersion measure
(DM) variations over time. Furthermore, we present the extreme
variations in the orbital period and projected semi-major axis of the
system, which are much larger than the ones published by
\cite{dlk+01}. In Section~\ref{sec:Disc} we rule out several possible
contributions to the aforementioned variations and we discuss the
physical mechanisms possibly responsible for those. Specifically, we
show that gravitational quadrupole coupling (GQC) and classical
spin-orbit coupling (SOC) can be viable mechanisms for the orbital
variations of the \psr\ system, under certain assumptions. In light of
the new orbital variation measurements, we discuss the possible
scenarios for the nature of the companion and the prospects for
observations of the system at higher energies, also in
Section~\ref{sec:Disc}. Finally, in Section~\ref{sec:conclusion}, we
briefly summarise our findings.

\section{Observations and Data Analysis}\label{sec:obs}
Description of the observing systems and the data acquisition
procedure of the EPTA telescopes can be found in \cite{lwj+09}, while
for Parkes this information is given in \cite{sbm+98}. A detailed
description on combining pulse times-of-arrival (TOAs) from different
telescopes can be found in \cite{jsk+08b,lwj+09}; details specific to
the data analysis used here, are listed below.

In our timing of \psr, all TOAs with uncertainties greater than
$20\,\mu$s were excluded as they don't impact the weighted
fit. Furthermore, 573 TOAs that were taken at orbital phases between
0.2 and 0.35 were not used because the potential excess column density
in the eclipsing region may cause additional modulation on the TOAs
derived from these data \citep{sbm+98,sbl+01a}.

Timing was performed with the \textsc{Tempo2} software package
\citep{hem06}, using the DE405 Solar-System ephemerides of the Jet
Propulsion Laboratory \citep{sta98b,sta04b} and the ELL1 binary model
\citep{lcw+01}. \textsc{Tempo2} minimises the weighted sum of the
squared residuals, producing a set of improved pulsar parameters and
post-fit timing residuals. In order to propagate unmodelled noise
through into the uncertainties of the timing model parameters, the TOA
uncertainties were multiplied by telescope-specific scaling factors to
achieve a reduced $\chi^2$ of unity. These scaling factors all fell
between 1.4 and 3.5, where the largest factor was applied to the TOAs
of the Effelsberg-Berkeley-Pulsar-Processor (EBPP). The large reduced
$\chi^2$ for the EBPP data is most likely caused by the fact that this
instrument only records total power (no polarisation information) and
has no RFI mitigation applied to it. Arbitrary phase offsets (also
called ``jumps'') were introduced between data obtained at different
frequencies and/or taken with different telescopes, to account for
frequency-dependent pulse shape evolution and observatory-dependent
differences in instrumental delays, cable lengths and geodetic
position, amongst others. These jumps were included as free parameters
in the least-squares fit performed by \textsc{Tempo2} and they absorb
any non-changing dispersive effects in our data. Therefore we were
unable to evaluate the average DM and hence fixed the value obtained
by \citet{sbm+98} as a constant in our timing model. However, the data
does contain some sensitivity to time-varying dispersive delays, which
is fully discussed in Section~\ref{subsec:DM}.

\begin{figure}
\begin{center}
  \includegraphics[width=0.45\textwidth, angle=0]{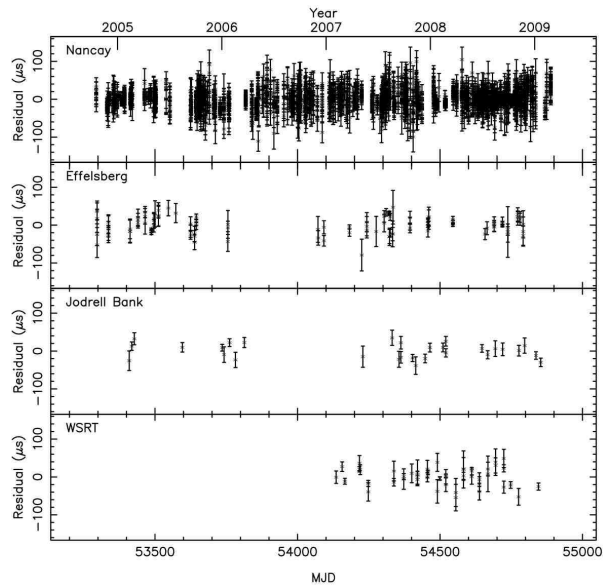}
  \caption{Post-fit timing residuals for the last 4.5 years of data. Our
    timing solution is given in Table~\ref{tab:par}. The majority of
    the data are from the Nan\c cay radio observatory, but data from
    Effelsberg, Jodrell Bank and Parkes were also used in the
    preceding nine years, which are not shown.}
\label{fig:postfit}
\end{center}
\end{figure}

In Table~\ref{tab:telescopes} the properties of the data from the
different telescopes are presented. Combination of the TOAs from all
the telescopes provides us with a 13-year dataset with no significant
gaps. In order to obtain a coherent timing solution for this combined
dataset, higher order derivatives (up to sixth order) of the orbital
period and projected semi-major axis must be fitted, leading to high
correlations between the modelled parameters. These variations in
orbital period and size are discussed in detail in sections
\ref{subsubsec:period} and \ref{subsubsec:xdot}. To avoid these
correlations and because 85\% of our data (and all of our
highest-precision data) were taken after 2004, we derive the timing
model, presented in Table~\ref{tab:par}, based solely on the last
4.5\,years of data, between MJDs 53293 and 54888 (as shown in
Figure~\ref{fig:postfit}). The remaining older data were exclusively
used to investigate potential variations in interstellar dispersion
(see Section~\ref{subsec:DM}) and to investigate possible causes for
the variations in orbital period and projected semi-major axis
(Section~\ref{subsec:orbit}).

\begin{table*}
\center
\caption{Properties of the individual telescope datasets.}
\begin{tabular}{lccccc}
\hline
    {Properties}     &{Effelsberg} & {Jodrell Bank}  & {Westerbork} &
    {Nan\c cay}  & {Parkes} \\
\noalign{\smallskip}
\hline
\noalign{\smallskip}
Number of TOAs                 &  490        & 139          & 46          & 2954 & 51\\
Time span (MJD)                & 50460--54791 & 49989--54853 & 54135--54845 & 53293--54888 & 49982--50343\\
Observed frequencies (MHz)     &  860, 1400, 2700 & 410, 606, 1400 &
330, 370, 840, 1380 & 1400 & 1400, 1700 \\
Typical integration (min) & 10 & 25 & 15 & 2 & 30\\
\noalign{\smallskip}
\hline
\end{tabular}
\label{tab:telescopes}
\end{table*}

\begin{table}
\caption{Timing parameters for \psr\ for the last epoch of observations 
(MJD range 53293--54888).}
\begin{tabular}{ll}
\hline
{Parameters} & {EPTA} \\
\noalign{\smallskip}
\hline
\noalign{\smallskip}
Right ascension, $\alpha$ (J2000) & 20$^{\rm h}$51$^{\rm m}$07$^{\rm s}$.51808(2)  \\
Declination,     $\delta$ (J2000) & $-$08$^\circ$27$^\prime$37$\farcs$7608(9)  \\ 
$\mu_{\alpha}$ (mas\,yr$^{-1})$   & 6.6(2)     \\
$\mu_{\delta}$ (mas\,yr$^{-1})$   & 3.2(7) \\
\\
$\nu$ (Hz)              & 221.796283737706(1)\\
$\dot{\nu}$ (s$^{-2}$)  & $-$6.2639(9)$\times 10^{-16}$  \\
$P$ (ms)                & 4.50864181828489(2)               \\ 
$\dot{P}$ (s\,s$^{-1}$) &  1.2733(2)$\times 10^{-20}$    \\  
\\
Reference epoch (MJD) & 54091  \\ 
MJD range of global timing model & 53293 -- 54888\\
MJD range epoch 1 & 49989 -- 51545 \\
MJD range epoch 2 & 51384 -- 52332 \\
MJD range epoch 3 & 51967 -- 53117 \\
MJD range epoch 4 & 52790 -- 54248 \\
MJD range epoch 5 & 54156 -- 54888 \\
\\ 
Dispersion measure, DM (cm$^{-3}$\,pc) $^{\ast}$ & 20.7458(2) \\
\\ 
Orbital period, $P_{\rm b}$ (days)  & 0.09911024846(2) \\ 
Projected semi-major axis, $x$ (lt-s) & 0.0450725(4)  \\ 
$\eta$   ($\equiv e\sin\omega$) & 5(1)$\times 10^{-5}$  \\ 
$\kappa$ ($\equiv e\cos\omega$) & 4(1)$\times 10^{-5}$  \\
Eccentricity, $e$ $^\ast$ & 6(1)$\times 10^{-5}$  \\
Longitude of the periastron, $\omega$ (deg) $^\ast$  & 52(12) \\
$T_{\rm ASC}$ (MJD) & 54091.0343503(1)  \\ 
\\ 
%
$\dot{P_{\rm b}}$, epoch 1 (s\,s$^{-1}$) &  $-1.33(6)\times 10^{-11}$\\
$\dot{P_{\rm b}}$, epoch 2 (s\,s$^{-1}$) &  $0.9(1)\times 10^{-11}$\\
$\dot{P_{\rm b}}$, epoch 3 (s\,s$^{-1}$) &  $1.8(3)\times 10^{-11}$\\
$\dot{P_{\rm b}}$, epoch 4 (s\,s$^{-1}$) &  $-1.81(3)\times 10^{-11}$\\
$\dot{P_{\rm b}}$, epoch 5 (s\,s$^{-1}$) &  $1.34(6)\times 10^{-11}$\\
$\dot{x}$, epoch 1 (s\,s$^{-1}$) &  $-1.7(4)\times10^{-13}$ \\
$\dot{x}$, epoch 2 (s\,s$^{-1}$) &  $-7.9(8)\times10^{-13}$ \\
$\dot{x}$, epoch 3 (s\,s$^{-1}$) &  $9(1)\times10^{-13}$ \\
$\dot{x}$, epoch 4 (s\,s$^{-1}$) &  $-0.8(2)\times10^{-13}$ \\
$\dot{x}$, epoch 5 (s\,s$^{-1}$) &  $0.2(2)\times10^{-13}$ \\
\\ 
Solar system ephemeris model & DE405  \\ 
Number of TOAs               & 3126    \\ 
RMS timing residual ($\mu$s) & 12.2      \\ 
\noalign{\smallskip} \hline
\end{tabular}
\\
$^\ast$ The eccentricity and the longitude of the periastron are
calculated from the Laplace-Lagrange parameters, $\eta$ and
$\kappa$. The DM value and its uncertainty were taken from
\citet{sbm+98}\\
Figures in parentheses are the nominal 1\,$\sigma$ \textsc{TEMPO2}
uncertainties in the least-significant digits quoted.\\
These parameters were determined with \textsc{Tempo2}, which uses the
International Celestial Reference System and Barycentric Coordinate
Time. Refer to \citet{hem06} for information on modifying this timing
model for observing systems that use \textsc{Tempo} format parameters.
\label{tab:par}
\end{table}
 
\section{Timing Results}\label{sec:timing}
\subsection{Proper motion}\label{subsubsec:pm}
From the data shown in Figure~\ref{fig:postfit}, we derived the first
significant measurement of proper motion in both right ascension and
declination. Using the NE2001 model for the Galactic distribution of
free electrons \citep{cl02} and the pulsar's dispersion measure (DM)
of $20.745$\,cm$^{-3}$\,pc, a distance of $d \simeq 1040$\,pc has been
derived. Combining the latter with the composite proper motion, $\mu_t
= 7.3 \pm 0.4\,$mas\,yr$^{-1}$, the velocity of the system can be
calculated:
\begin{equation}
\label{eq:velocity1}
v_t = \mu_t d = 36.1 \pm 7.5\,{\rm km\,s}^{-1},
\end{equation}
being consistent with previous estimates \citep{sbm+98}. The largest
part of the error originates from the DM-derived distance where a 20\%
uncertainty was assumed.

\subsection{Orbital eccentricity}\label{subsubsec:ecc}
\psr\ is in a low-eccentricity binary. A precise eccentricity
measurement has been challenging in the past but from our timing one
has been made for the first time: $e = (6.2 \pm 1.3) \times
10^{-5}$. This eccentricity value is much larger than expected for
such a tight binary system if plotted in an eccentricity versus
orbital period diagram \citep[as in][]{phi92,pk94}. Although the model
curves do not extend to very low orbital periods, this still
constitutes evidence that the \psr\ system at its present evolutionary
stage experiences processes that do not allow for perfect
circularisation. These processes could very well be density
fluctuations in a convective envelope of the donor star (see
Section~\ref{subsubsec:applegate}). Hence the eccentricity of
\psr\ ($6.2 \times 10^{-5}$) is much larger than the expected residual
eccentricities ($\leq 10^{-8}$) for binary millisecond pulsars with a
similar orbital period \citep{lr01}.

\subsection{DM variations}\label{subsec:DM}

As described in Section~\ref{sec:Intro}, low-frequency observations of
this system show eclipses as the pulsar passes through
periastron. This may imply increased dispersion as a function of
orbital phase, which could corrupt measurements of binary
parameters. To investigate this possibility, we took the high-quality,
multi-frequency data from the final 4.5\,years and measured the DM as
a function of orbital phase while keeping the jumps between observing
bands and observatories fixed. This did not result in significant
trends at any orbital phase, including egress or ingress\footnote{This
  does not contradict the increased column density in the eclipse
  region but underlines the limited sensitivity of our dataset to DM
  variations.}. Since this experiment required a fit for DM over TOAs
that were restricted to a small fraction of the binary phase, a
simultaneous fit to standard orbital parameters was
impossible. However, the wide frequency range available at any given
orbital phase implies that even if orbital-phase--dependent DM
variations did occur, their effect on the TOAs could only partially be
absorbed by the orbital parameters, implying some residual DM
variations would still be visible. Simultaneous fitting for DM and
long-term effects (such as pulsar position and proper motion, spin,
spin-down and orbital derivatives $\dot{P}_{\rm b}$ and $\dot{x}$)
over limited orbital-phase ranges did not affect the DM estimates or
the other timing parameters, confirming that the long-period terms of
the timing model are not affected by any potential
orbital-phase--dependent variations in interstellar dispersion.

\begin{figure}
\begin{center}
  \mbox{\includegraphics[width=0.34\textwidth, angle=-90]{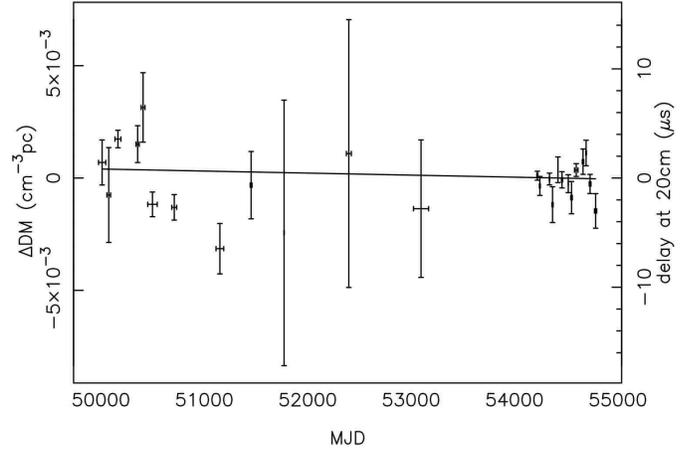}}
  \caption{DM variations versus time, including the best-fit linear
    trend.}
\label{fig:DMevol}
\end{center}
\end{figure}

In addition to investigating the DM evolution as a function of orbital
phase, the DM variations were determined across the entire data
set. To this end, intervals with adequate multi-frequency data were
identified by hand and DM values were fitted in each interval, while
jumps between telescopes and observing bands were kept fixed at the
values determined from the entire data set. Clearly this cannot result
in accurate measurements of DM, but it does allow variations to be
determined precisely. The resulting measurements (shown in
Fig.~\ref{fig:DMevol}) demonstrate that no complex DM evolution is
present in this data set, though a shallow DM decay is measured: ${\rm
  d}({\rm DM})/{\rm d}t = \left(-4.3 \pm 1.4\right) \times
10^{-4}\,{\rm cm}^{-3}\,{\rm yr}^{-1}\,{\rm pc}$. This measurement is
barely significant at the 3-$\sigma$ level. Given the small number of
high-quality DM estimates that contributed to it and the possible
underestimation of the DM measurement uncertainties (e.g. variations
in orbital parameters could have affected the offset between
non-simultaneous observations), we did not include a time-dependent DM
variation in our timing model.

\subsection{Companion mass}
\begin{figure}
\begin{center}
  \mbox{\includegraphics[width=0.34\textwidth,angle=-90]{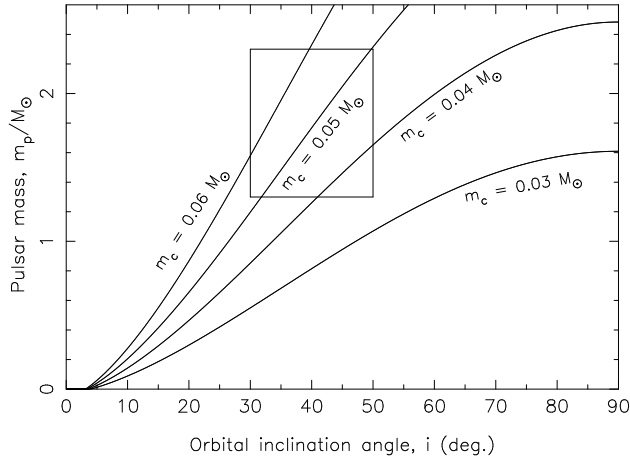}}
  \caption{The masses of the stellar components in the J2051$-$0827
    system as a function of orbital inclination angle. The box
    indicates the most likely parameter space in terms of the neutron
    star mass and constraints from radio eclipses and optical
    observations (see text).}
\label{fig:mass}
\end{center}
\end{figure}

Optical observations \citep{svbk01b} yielded a best fit for the
orbital inclination angle of about 40$^{\circ}$. Depending on the
unknown mass of the pulsar this results in a companion mass in the
range: $m_{\rm c}\simeq 0.04 - 0.06{\rm \,M}_{\odot}$ (see
Figure~\ref{fig:mass}) given the constraints from the mass function:
\begin{equation}
  \label{mfunc}
  f = \frac{(m_{\rm c} \sin i)^3}{(m_{\rm p}+m_{\rm c})^2}
  =\frac{4\pi^2}{G}\,\frac{(a_{\rm p} \sin i)^3}{P_{\rm b}^2} =
  1.0030\times 10^{-5}\,M_{\odot},
\end{equation} 
where the measured observables are the orbital period, $P_{\rm b}$ and
the projected semi-major axis of the pulsar orbit, $a_{\rm p} \sin
i$. For the rest of this paper we shall assume a companion star mass
of $m_{\rm c} = 0.05{\rm \,M}_{\odot}$ and consequently a pulsar mass
of $m_{\rm p} = 1.8{\rm \,M}_{\odot}$. The pulsar mass of \psr\ is
likely to be significantly larger than the typical neutron star mass
($1.35{\rm \,M}_{\odot}$) obtained from measurements in double neutron
star binaries, given that this system has evolved through a long
($\sim$ Gyr) LMXB phase with sub-Eddington mass transfer \citep[see,
  e.g.][]{prp02}. Recent work on the mass determination of the
original black widow pulsar (PSR~B1957+20) by \citet{vbk11} confirms
that the pulsars in these systems can accrete significant amounts of
matter. For a discussion of the effect of irradiation on the accretion
efficiency in LMXBs, see \citet{rit08}.  The Roche-lobe radius of the
companion star in \psr\ changes slightly with the estimated stellar
masses and is found to be $R_{\rm L} = 0.15 \pm 0.01{\rm
  \,R}_{\odot}$. However, the size (radius) of the irradiated
companion star is difficult to determine accurately and in the
discussion further on we shall assume two different values for the
filling factor (the ratio of the volume-equivalent radius of the
companion star to its Roche lobe) of 0.43 and 0.95 \citep{svbk01b},
corresponding to stellar radii of about $0.064\,$R$_{\odot}$ and
$0.14\,$R$_{\odot}$ respectively.

\subsection{Orbital changes}\label{subsec:orbit}
In order to monitor their variations over time, values for $P_{\rm b}$
and $x$ were derived for each year of data, where three months of
overlap were kept between adjacent years. In doing so, all model
parameters besides $x$, $P_{\rm b}$ and $T_{\rm ASC}$ were held fixed
and the timing reference epoch was defined to be the centre of each
year-long interval. 
The fractional changes of these measurements 
are shown in Figure~\ref{fig:period}. Clearly, five different epochs
can be identified, in which variations of both $P_{\rm b}$ and $x$ can
be described using only a linear trend, as shown in the
figure.

\section{Discussion}\label{sec:Disc}
\subsection{Orbital period variations}\label{subsubsec:period}

\begin{figure}
\begin{center}
  \includegraphics[width=0.48\textwidth]{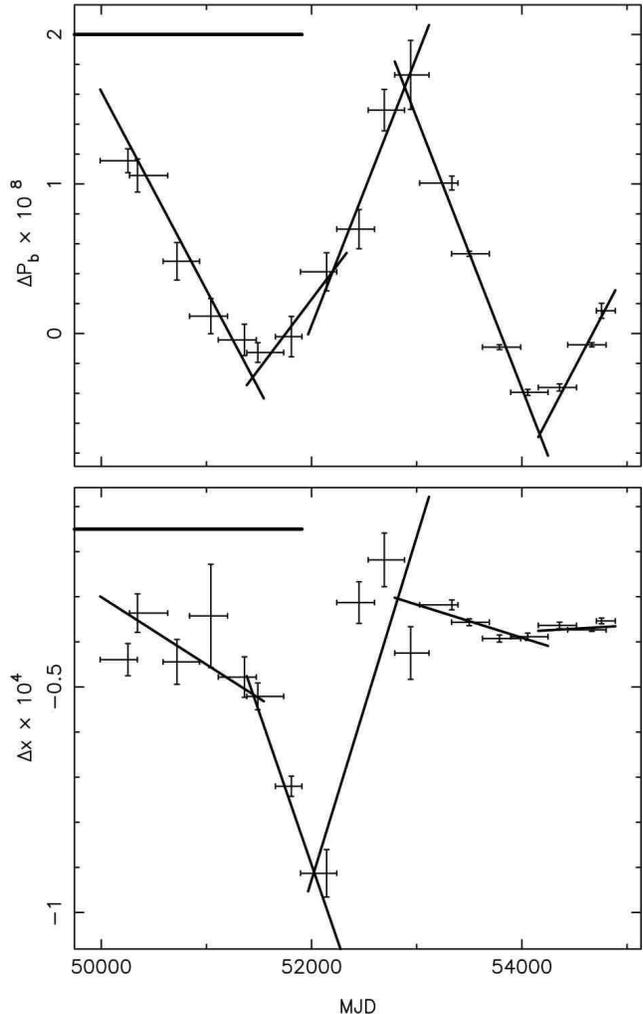}
  \caption{Changes of the orbital period (top) and projected
    semi-major axis (bottom) versus date. Each point corresponds to
    close to one year of data, with typically three months overlap
    between adjacent points, though the precise amount of overlap
    varies depending on the density of observations and the amplitude
    of the orbital variations. The horizontal bars indicate the
    intervals over which measurements were made, and the uncertainties
    of the measurements are shown by vertical error bars that are
    placed at the mean TOA of the data contained in the
    interval. Because of inhomogeneous sampling, the mean TOA is not
    necessarily at the middle of the interval. For each of the five
    epochs listed in Table~\ref{tab:par} a trend line derived from the
    respective $\dot{P}_{\rm b}$ and $\dot{x}$ measurements given in
    Table~\ref{tab:par} is shown. In most cases these trend lines fit
    well through the measurement points, though deviations may be
    caused by rapid variations in orbital parameters, underestimation
    of measurement uncertainties, correlations between parameters and
    the fact that on many of the shorter measurement intervals only
    part of the timing model could be fitted for, incurring potential
    corruptions in correlated parameters. The thick horizontal bars on
    the top of the figure indicate the timing baseline of
    \citet{dlk+01}, which only really spans the first two of these
    epochs.}
\label{fig:period}
\end{center}
\end{figure} 

$\dot{P_{\rm b}}$ is the observable rate of change of the orbital
period and is caused by a variety of effects, both intrinsic to the
system and caused by kinematic effects relative to the observer. The
most important contributions are:
\begin{equation}
  \label{eq:period}
  \dot{P_{\rm b}} = \dot{P_{\rm b}}^{\rm GW} + \dot{P_{\rm b}}^{\rm D} +
  \dot{P_{\rm b}}^{\dot{m}} + \dot{P_{\rm b}}^{\rm T} + \dot{P_{\rm b}}^{\rm Q}. 
\end{equation}
As a point of reference, the values for $\dot{P}_{\rm b}$ in the two
most extreme epochs are $\dot{P_{\rm b}} = -1.81(3) \times 10^{-11}$
and $\dot{P_{\rm b}} = 1.8(3) \times 10^{-11}$.

The first term, $\dot{P_{\rm b}}^{\rm GW}$, is the contribution due to
gravitational wave emission. In general relativity, for circular
orbits it is given by \citep{pet64}:
\begin{equation}\label{eq:gr}
  \dot{P}_{\rm b}^{\rm GW} = -\frac{192\pi}{5}
  \left( \frac{2\pi}{P_{\rm b}}\,\frac{Gm_{\rm c}}{c^3} \right)^{5/3}
    \frac{q}{(q+1)^{1/3}},
\end{equation}
The mass ratio $q = 36$ has been calculated assuming a pulsar mass $m_{\rm
 p}=1.8$\,M$_{\odot}$ and a companion mass $m_{\rm c}=0.05$\,M$_{\odot}$ for
an inclination angle of $i=40^{\circ}$.  For \psr\ we find $\dot{P_{\rm
 b}}^{\rm GW} \simeq -7.5 \times 10^{-14}$. This value is about three
orders of magnitude less than the observed value of $\dot{P_{\rm b}}$.

The second term, $\dot{P_{\rm b}}^{\rm D}$, is the Doppler correction,
which is the combined effect of the proper motion of the system
\citep{shk70} and a correction term for the Galactic acceleration. The
contribution for the Galactic acceleration at the location of \psr,
$\dot{P_{\rm b}}^{\rm Gal}$, is of order $1.1 \times 10^{-15}$
\citep{lwj+09}. Using numbers from Table~\ref{tab:par}, we also
calculate the contribution due to the Shklovskii effect according to
the following:
\begin{equation}
\label{eq:shk}
\dot{P_{\rm b}}^{\rm Shk} = \frac{(\mu_{\alpha}^{2} +
  \mu_{\delta}^{2})d}{c}\,P_{\rm b} 
  \simeq 1.1 \times 10^{-15}.
\end{equation}
By summing we yield the Doppler correction:
\begin{equation}
\label{eq:doppler}
\dot{P}_{\rm b}^{\rm D} = \dot{P}_{\rm b}^{\rm Gal} + \dot{P_{\rm
    b}}^{\rm Shk} \simeq 2.2 \times 10^{-15}, 
\end{equation}
four orders of magnitude smaller than the measured value. 

An acceleration of the binary system with respect to the Solar System
Barycentre (SSB) could also be caused by a third massive body orbiting
the binary system. However, it would affect the orbital period
derivative and the spin period derivative in the same way. Assuming
that the spin period derivative is caused entirely by the
acceleration, we can estimate the maximal effect this would have on
$\dot{P}_{\rm b}$: $(\dot{P}_{\rm b}/P_{\rm b})^{\rm
  acc}=(\dot{P}/P)\simeq -3 \times 10^{-18}$\,s$^{-1}$. This is seven
orders of magnitude smaller than the measured value, thus there can be
no massive third body orbiting the system.

The third term, $\dot{P_{\rm b}}^{\dot{m}}$, is the contribution from
the mass loss of the binary. If we consider a circular orbit, no mass
loss from the pulsar ($\dot{m_{\rm p}} = 0$) and a rate of mass loss
from the companion $\dot{m_{\rm c}} > 0$ \citep{jea24}, we obtain:
\begin{equation}
\label{eq:mass}
  \frac{\dot{P}_{\rm b}}{P_{\rm b}} = -2 \frac{\dot{m}_{\rm c}}{M}
\end{equation}
and therefore:
\begin{equation}
\dot{m}_{\rm c} 
   = -\frac{M}{2}\,\frac{\dot{P}_{\rm b}}{P_{\rm b}} 
   \simeq 6.2 \times 10^{-8} \,{\rm M_{\odot}\,yr^{-1}}, 
\end{equation}
where $M$ is the total mass of the system. This number is six orders
of magnitude larger than expected \citep[as calculated
  by][]{sbl+96} and hard to reconcile with the low electron densities
measured in the eclipse region \citep{sbl+96}. In addition, for the
second of the epochs we would need mass injection to the companion in
order to explain the orbital decay -- an infeasible scenario for this
system.

Tidal forces in tight synchronous binaries may also be responsible for
interactions which would change the orbital period. This could be the
case if, for example, a companion star suddenly contracts or expands
and thereby changes its spin angular momentum. Tidal locking of the
orbit would then result in exchange of spin and orbital angular
momentum, e.g. \citet{ts01}. However, in \psr\ we do not see reasons
for any sudden major radial changes in the structure of the companion
star -- especially not changes that would not result in detection of
{X}-ray bursts (A. Levine and D. Altamirano, private communication).
Hence, we neglect the term $\dot{P_{\rm b}}^{\rm T}$.

Since all the above contributions are much smaller than the observed
variation of the orbital period, we conclude that they must originate
from the last term of Equation~(\ref{eq:period}), which represents
variations of the gravitational quadrupole moment of the companion star.

\subsubsection{The Applegate GQC model}\label{subsubsec:applegate}
The gravitational quadrupole coupling (GQC) mechanism \citep{app92,
  as94} has been applied successfully to the eclipsing binary system
PSR~B1957$+$20 \citep{as94}, explaining the orbital period variations
seen by \cite{aft94}. In addition, it was proposed by \cite{dlk+01} as
the main mechanism for producing the orbital period variations of
\psr. However, as we will present below, the latter publication
considered only a small part of these variations, which led to an
underestimation in their calculations.

In the GQC model \citep{as94}, magnetic activity is driven by energy
flows in convective layers of the irradiated companion. This, in
combination with wind mass loss, results in a torque on its spin,
which holds it slightly out of synchronous rotation, causing tidal
dissipation of energy and heating of the companion. The resultant
time-dependent gravitational quadrupole moment (e.g. variations of the
oblateness) causes modulation of the orbital period on a short, in
principle dynamical, time scale. An increase in the quadrupole moment
causes the two stars to move closer together and a decrease in the
quadrupole moment results in a widening of the orbit -- in Applegate's
model total orbital angular momentum is assumed to remain constant and
the neutron star is treated as a point mass. The details of the
hydrodynamic dynamo and its activity cycle remain unspecified for our
purposes (also given the unknown nature of the companion star -- see
Section~\ref{subsec:companion}) and it is simply assumed that the
variable quadrupole moment, $\Delta Q$, is caused by cyclic spin-up
and spin-down of the outer layers of the companion, which leads to
orbital period changes equal to \citep{as94}:
\begin{equation}\label{eq:dqdp}
  \left(\frac{\Delta{P}_{\rm b}}{{P_{\rm b}}}\right)^Q =
  -9\frac{\Delta{Q}}{m_{\rm c}a^2},
\end{equation}
where $a$ is the separation and $m_{\rm c}$ the companion mass. The
transfer of angular momentum $\Delta J$ to a thin shell of radius
$R_{\rm c}$ and mass $M_{\rm s}$, rotating with angular velocity
$\Omega$ in the gravitational field of a point mass $m_{\rm c},$ will
cause the quadrupole moment of the shell to change by \citep{as94}:
\begin{equation}
\label{eq:dq}
\Delta Q =\frac{2}{9} \, \frac{M_{\rm s} R_{\rm c}^5}{Gm_{\rm c}} \,
          \Omega \, \Delta\Omega \;,
\end{equation}
where $G$ is the Newtonian gravitational constant and $\Delta\Omega$
is the change in the angular velocity of the shell. In order to
produce orbital period changes $\Delta P_{\rm b}$ you need a variable
angular velocity given by
\begin{equation}
\label{eq:dOMEGA}
\frac{M_{\rm s}}{m_{\rm c}}\frac{\Delta \Omega}{\Omega} =\frac{Gm_{\rm
    c}}{2R_{\rm c}^3}\left(\frac{a}{R_{\rm
      c}}\right)^2\left(\frac{P_{\rm b}}{2\pi}\right)^2\frac{\Delta
  P_{\rm b}}{P_{\rm b}}
\end{equation}
\citep{as94}. This procedure produces a luminosity variation $\Delta
L$ given by
\begin{equation}
\label{eq:dL}
\Delta L =\frac{\pi}{3}\frac{Gm_{\rm c}^2}{R_{\rm c}P_{\rm
    mod}}\left(\frac{a}{R_{\rm c}}\right)^2\frac{\Omega_{\rm
    dr}}{\Omega}\frac{\Delta P_{\rm b}}{P_{\rm b}},
\end{equation}
where $P_{\rm mod}$ is the period of the orbital period modulation and
$\Omega_{\rm dr}$ is the angular velocity of the differential
rotation \citep{as94}.

\cite{as94} assume in their model a mass $M_{\rm s}\simeq 0.1 m_{\rm
  c}$ for the thin shell; an angular velocity $\Omega_{\rm dr} \simeq
\Delta \Omega$; and luminosity variations at the $\Delta L/L \simeq
0.1$ level. From Figure~\ref{fig:period} we calculated for \psr\ a
total change in the orbital period of $\Delta P_{\rm b}/P_{\rm b}
\simeq 2.2 \times 10^{-7}$ and a modulation period of $P_{\rm mod}
\simeq 7.5$\,yr. As mentioned above these two values where under- and
over-estimated, respectively, in \cite{dlk+01} because of their
significantly smaller data set. By assuming an inclination angle of $i
= 40^{\circ}$ and a companion radius of $R_{\rm c}\sim
0.064$\,R$_{\odot}$ \citep{svbk01b} for a separation $a \simeq 7.1
\times 10^8$\,m and a companion mass $m_{\rm c} = 0.05$\,M$_\odot$, we
derive a variation of the angular velocity of the companion of:
\begin{equation}
\label{eq:domeq}
\frac{\Delta \Omega}{\Omega} \sim 4\times 10^{-2}.
\end{equation}
This variation would produce a variable luminosity of $\Delta L\sim
1.4 \times 10^{33}$erg\,s$^{-1}$ and according to the model the
internal luminosity of the companion must be
\begin{equation}\label{eq:luminosity}
  L \sim 1.4 \times 10^{34}\,{\rm erg}\,{\rm s}^{-1}.
\end{equation}
From the optical observations of the companion \citep{svbk01b} we can
estimate a maximum effective temperature of $T_{\rm eff}^{\rm
  max}\simeq 3000$\,K which gives an internal luminosity of $L \sim
10^{30}$\,erg\,s$^{-1}$. We therefore conclude that, under the
assumptions listed above, the GQC model produces much larger orbital
period variations than expected and cannot fit the \psr\ system. If
the companion was tidally powered, the internal luminosity derived
from the optical observations would be much higher. One reason why GQC
does not fit the observations may be the simplifications of the model
itself, i.e. the consideration that all of the energy variations
appear as luminosity variations without any loss
\citep{app92}. Although improvement of the model could possibly
decrease $\Delta L$ by a factor of a few, the difference we calculate
is significantly larger.

The previous calculations can change by a large factor if we consider
the alternative model that \citet{svbk01b} presented for the companion
star, where the latter almost fills its Roche lobe ($R_{\rm c} =
0.14\,$R$_{\odot}$). Using the values for that model, we derive a
variation of the angular velocity of the companion of
\begin{equation}
\label{eq:domeq2}
\frac{\Delta \Omega}{\Omega} \sim 7.1\times 10^{-4}
\end{equation}
and an internal luminosity of 
\begin{equation}
\label{eq:luminosity2}
L \sim 2 \times 10^{31}\,{\rm erg\,s}^{-1}.
\end{equation}
This number is much closer to the newly calculated internal luminosity
from the optical of $L \sim 6\times 10^{30}$\,erg\,s$^{-1}$. Thus,
under specific assumptions the GQC model can explain the orbital
period variations of \psr.

\subsection{Changes in the projected semi-major axis}\label{subsubsec:xdot}

The observed values of $\dot{x}$ can be the result of various effects
\citep{lk05}:
\begin{equation}
\label{eq:dotA1}
\dot{x}_{\rm obs} = \dot{x}^{\rm GW} + \dot{x}^{\rm D} + \frac{{\rm d}\epsilon_{\rm
    A}}{{\rm d}t} + \dot{x}^{\rm PM} + \dot{x}^{\dot{m}} + \dot{x}^{\rm Q} +
\dot{x}^{\rm SOC}.
\end{equation}
For the most extreme epochs, we have: $\dot{x} = -7.9(8)\times 10^{-13}$
and $\dot{x} = 9(1) \times 10^{-13}$.

The first term, $\dot{x}^{\rm GW}$ arises from orbital shrinkage due to
gravitational-wave damping. Using Kepler's third law and Equation~\ref{eq:gr}:
\begin{equation}
\label{eq:xdotGR}
\frac{\dot{x}^{\rm GW}}{x} = \frac{2}{3}\frac{\dot{P}_{\rm b}^{\rm
    GW}}{P_{\rm b}} \simeq -2.6 \times 10^{-19}
\end{equation}
\citep{pet64}. This contribution is much smaller than the current
measurement.

The second term, $\dot{x}^{\rm D}$, is identical to the second term of
Equation~(\ref{eq:period}). The contribution for the Galactic
acceleration is of order $1.1 \times 10^{-15}$ and the contribution of
the Shklovskii effect $\dot{x}^{\rm Shk} = x (\mu_{\alpha}^{2} +
\mu_{\delta}^{2}) d / c \sim 6.0 \times 10^{-21}$. Both of these are
very small compared to the observed value so this term can be
neglected.

The third term, ${\rm d}\epsilon_{\rm A}/{\rm d}t$, is the contribution of the
varying aberration caused by geodetic precession of the pulsar spin
axis and is typically of order $\Omega^{\rm geod}P/P_{\rm b} \approx
7.4 \times 10^{-17}$ \citep{dt92}. For a recycled pulsar, like \psr,
the spin is expected to be close to parallel to the orbital angular
momentum, which further suppresses this effect. Hence, the
contribution is at least three orders of magnitude smaller than the
observed value.

The fourth term, $\dot{x}^{\rm PM}$, represents a variation of $x$
caused by a change of the orbital inclination while the binary system
is moving relatively to the SSB \citep{ajrt96, kop96}. This effect is
quantified by the following equation:
\begin{equation}\label{eq:dotx}
  \dot{x}^{\rm PM} = 1.54\times 10^{-16}\,x\,\cot i\,
                     (-\mu_\alpha \sin\Omega_{\rm asc} 
                      +\mu_\delta \cos\Omega_{\rm asc}),
\end{equation}
where $\Omega_{\rm asc}$ is the position angle of the ascending node. The
quantities $x$, $\mu_\alpha$ and $\mu_\delta$ are expressed in seconds
and milliarcseconds per year, respectively. The maximal contribution
of the proper motion is:
\begin{equation}\label{eq:dotxpm}
  \dot{x}^{\rm PM}_{\rm max} = 1.54 \times 10^{-16}\,x\,
    (\mu_\alpha^2 + \mu_\delta^2)^{1/2} \, \cot i.
\end{equation}
For an inclination angle of $i = 40^{\circ}$, we get $\dot{x}^{\rm
  PM}_{\rm max} \simeq 6.0 \times 10^{-17}$. Thus, this term is also
very small compared to the measured $\dot{x}$.

The fifth term, $\dot{x}^{\dot{m}}$, represents a change in the size
of the orbit caused by mass loss from the binary system. Using
Equation~(\ref{eq:mass}) and Kepler's third law, we calculate the rate
of mass loss from the companion to be seven to eight orders of
magnitude larger than expected.

The sixth term, $\dot{x}^{\rm Q}$, is caused directly by the change in
quadrupole moment as described in
Section~\ref{subsubsec:period}. Because orbital angular momentum is
conserved in the Applegate model, a change in the orbital period is
related to a change in the size of the orbit by
\begin{equation}\label{eq:APb}
  \frac{\Delta a}{a} = 2 \frac{\Delta P_{\rm b}}{P_{\rm b}}.
\end{equation}
It can be seen from Figure~\ref{fig:period} that this contribution
falls short by four orders of magnitude to explain the observed
changes in the projected semi-major axis $x$.

As all other contributions are much smaller than the observed
variation of the projected semi-major axis, they must originate from
the last term of Equation~(\ref{eq:dotA1}): the classical spin-orbit
coupling (SOC) term. 

\subsubsection{The Applegate GQC model with SOC}\label{sssec:GQCSOC}
Through SOC, the quadrupole of a rapidly rotating companion leads to
apsidal motion and precession of the binary orbit. This in turn causes
a variation of the longitude of periastron (which is impossible to
measure in a system with such small eccentricity) and of the projected
semi-major axis, according to:
\begin{equation}
\label{eq:soc}
  \dot x^{\rm SOC} = 
    x\,n_{\rm b}\,\tilde Q\,\cot i\,\sin\theta\,\cos\theta\,\sin\Phi, 
\end{equation}
\citep{sb76, wex98}, where $n_{\rm b} = 2\pi/P_{\rm b}$ is the orbital
frequency, $\theta$ is the angle between the spin and orbital angular momentum
and $\Phi$ is the longitude of the ascending node with respect to the
invariable plane (plane perpendicular to the total angular momentum). The
dimensionless quadrupole $\tilde Q$ is related to the quadrupole $Q$ by
\begin{equation}\label{eq:Qnorm}
  \tilde Q = \frac{3}{2}\,J_2 \left(\frac{R_{\rm c}}{a}\right)^2 \;,
  \quad J_2 = \frac{3Q}{m_{\rm c}R_{\rm c}^2} \;,
\end{equation}
with $J_2$ the dimensionless measure of the quadrupole moment. Since
no uniform change in $x$ is detected, we can consider the
companion to consist of a long-term stable component with spin axis
aligned to the orbital angular momentum axis (i.e. $\theta \approx 0$)
and in addition to that, a part of the star (like e.g. an outer shell)
that changes its quadrupole moment by $\Delta \tilde{Q}$ and has an
effective and variable angle $\theta$ independent of the rest of the
star. This part will then give rise to a variation $\dot{x}/x$ of the
order $n_{\rm b} \Delta \tilde {Q}$.

In Section~\ref{subsubsec:applegate} we have described how changes in
the gravitational quadrupole moment of the companion could cause the
observed $P_{\rm b}$ variations. Even though these direct GQC effects
are insufficient to explain the $x$ variations (i.e. $\dot{x}^{\rm Q}
= 0$ as explained above), through the SOC mechanism they might have a
more substantial effect. 

Extending Equation~\ref{eq:dqdp} to arbitrary $\theta$ and using
Equation~\ref{eq:Qnorm}, we have:
\begin{equation}\label{eq:PbQtilde}
\frac{\Delta P_{\rm b}}{P_{\rm b}} = -2 \Delta \tilde{Q} \left( 1-
  \frac{3}{2} \sin^2{\theta}\right).
\end{equation}
Combining this with the observed $\dot{P}_{\rm b}$ values
(Table~\ref{tab:par}) and assuming $\theta = 0$ for now, we get
the maximum $\Delta \tilde{Q}$ values presented in
Table~\ref{tab:DQnot}. Comparison to the $\dot{x}/x$ values from
Figure~\ref{fig:period}, also listed in Table~\ref{tab:DQnot}, shows
that while the order of magnitude is rougly correct, the sign does
neither correlate nor anti-correlate, which means that the orientation
of the quadrupole-moment changes (i.e. $\theta$) must vary strongly
from epoch to epoch, which may be
unphysical.

\begin{table}
\center
\caption{Quadrupole moment variations for the five epochs, as derived
  from the $P_{\rm b}$ variations, alongside the $x$ variations as observed.}
\begin{tabular}{rrrr}
\hline
{Epoch} & $\Delta Q$ & $n_{\rm b} \Delta \tilde{Q}$ & $\dot{x}/x$ \\
&($10^{45}$\,g\,cm$^2$) & ($10^{-11}$\,s$^{-1}$) & ($10^{-11}$\,s$^{-1}$) \\
\noalign{\smallskip}
\hline
\noalign{\smallskip}
1    & $ 11.2$ &  $7.7$ & $-0.38$\\
2    & $ -4.6$ & $-3.2$ & $-1.8$\\
3    & $-11.3$ & $-7.7$ &  $2.0$\\
4    & $ 14.3$ &  $9.8$ & $-0.18$\\
5    & $ -5.3$ & $-3.6$ &  $0.044$\\
\noalign{\smallskip}
\hline
\end{tabular}
\label{tab:DQnot}
\end{table}

\begin{figure}
\begin{center}
  \mbox{\includegraphics[width=0.41\textwidth]{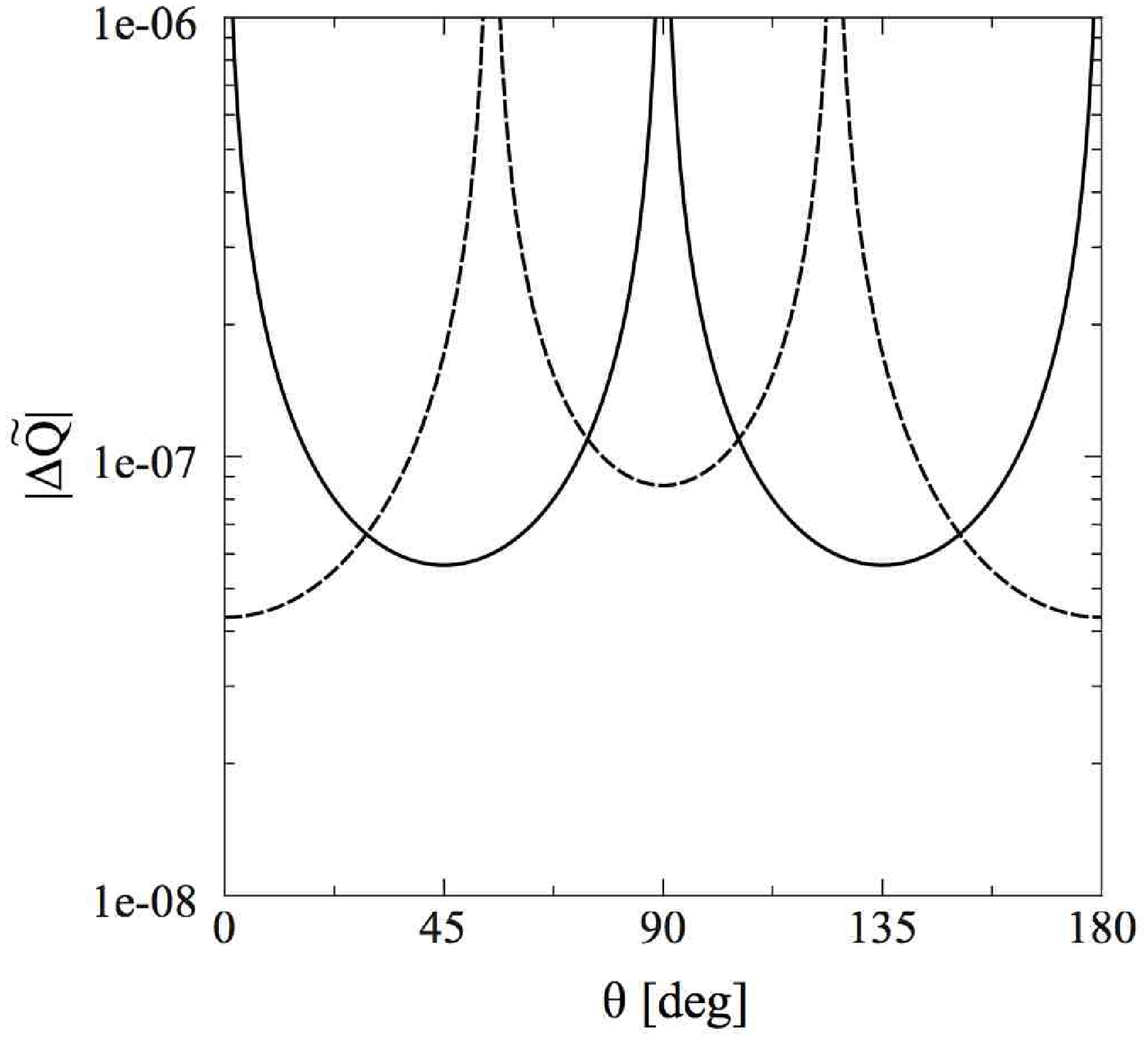}}
  \mbox{\includegraphics[width=0.41\textwidth]{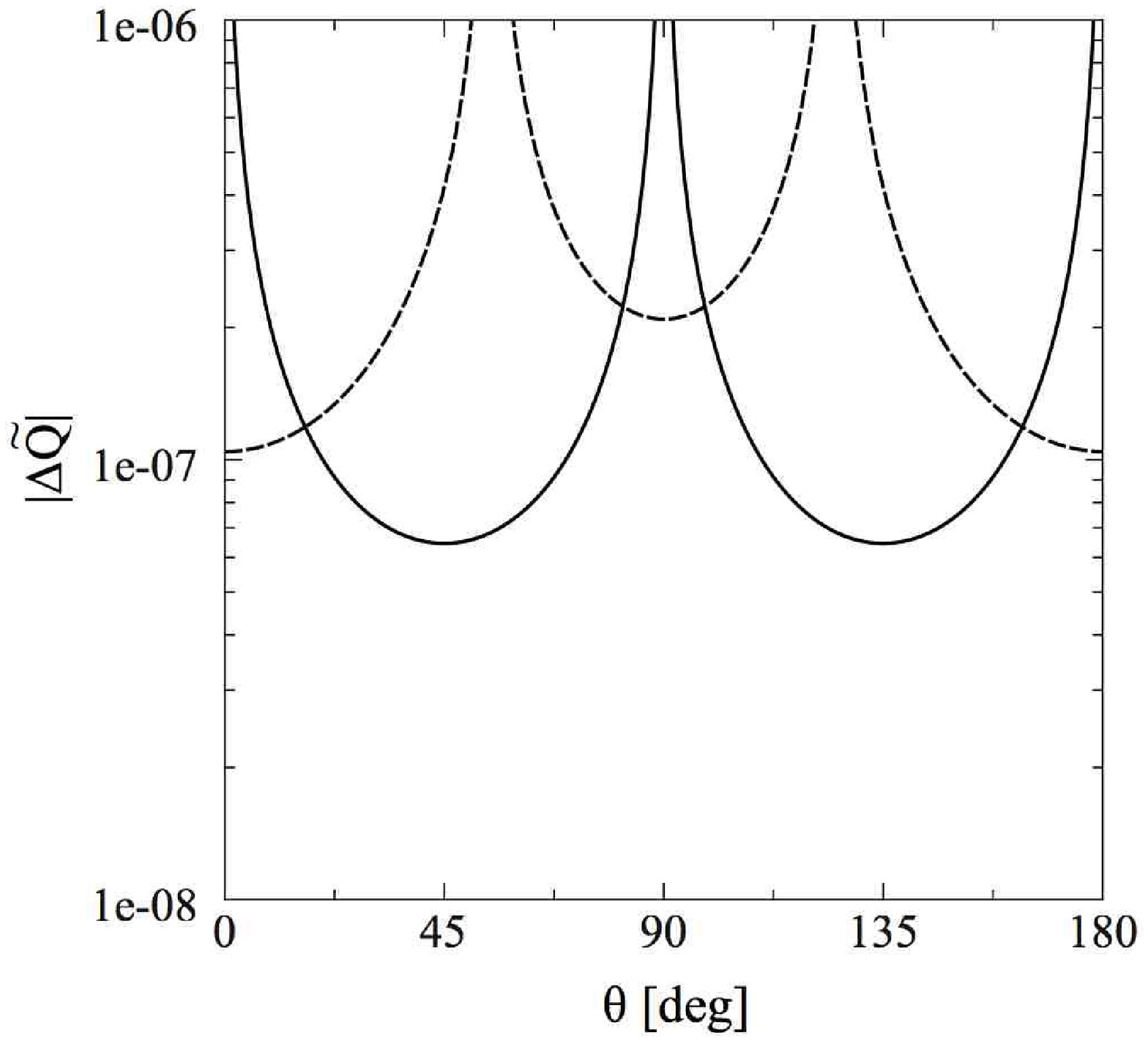}}
  \caption{Quadrupole moment changes versus $\theta$ as produced by
    SOC (solid line) and GQC (dashed line) for epochs two (top) and
    three (bottom).}
\label{fig:|DQ|}
\end{center}
\end{figure}

In order to examine more closely if the orbital variations can be
produced by GQC and SOC arising from the same $\Delta Q$, we have
plotted in Figure~\ref{fig:|DQ|} how $|\Delta \tilde{Q}|$ depends on
$\theta$, assuming $\Phi = 45^{\circ}$ and using $i = 40^{\circ}$. The
curves are plotted for SOC, as derived from Equation~\ref{eq:soc}, and
for GQC, as derived from Equation~\ref{eq:PbQtilde}. The intersection
of the two lines clearly shows that for several values of $\theta$ a
$\Delta \tilde Q$ could explain both the orbital period and projected
semi-major axis variations for \psr, though these $\theta$ values
change significantly with epoch and we do not see how this easily fits
with the physical mechanism behind the thin shell model of
\citet{app92} (see also our Section~\ref{subsubsec:applegate}).

\subsubsection{Spin precession of the companion star}
In Section~\ref{sssec:GQCSOC} we assumed a mostly stable companion
star with an outer shell responsible for the changing quadrupole
moment. An alternative explanation for the variations in $P_{\rm b}$
and $x$ could be provided by limited quadrupole changes (as needed for
the $P_{\rm b}$ varations, see Section~\ref{subsubsec:applegate})
combined with an overall small tilt $\theta$ of the entire companion
star with respect to the orbit. This tilt causes a precession of both
the star and the orbit and it allows the entire quadrupole moment of
the star $\tilde{Q}$ to cause variations in $x$ according to
Equation~\ref{eq:soc}. We stress that in this scenario there need not
be a physical mechanism relating the spin precession (which causes the
$x$ variations) to the quadrupole moment changes (that induce the
$P_{\rm b}$ changes)\footnote{Note the changes in $\theta$ will have
  some contribution to the $P_{\rm b}$ changes as well. Based on
  Equation~5 of \citet{app92} we determined this effect and for all
  epochs save the third, this effect is at or well below the few
  percent level of the $P_{\rm b}$ changes induced by $\Delta Q$. In
  the third epoch, which requires the largest angle $\theta$, it
  contributes 1\% of the observed $P_{\rm b}$ variations.}.

To evaluate this model, we estimate the quadrupole moment of the
companion star assuming a non-degenerate structure\footnote{Deviations
  of a semi-degenerate nature should not be important in this case.}
and following \citet{lr99b}:
\begin{equation}
\label{Qcomp}
Q =
\frac{5}{6\xi}\,\frac{R^3\,\left[\mathcal{T}-(1-3\eta)\mathcal{M}\right]}
{GM}
\end{equation}
where $R$ and $M$ are the radius and mass of the star, $\mathcal{T}$ is its
rotational kinetic energy ($\mathcal{T} = I\Omega^2/2$), $\mathcal{M}$ is its
total magnetic energy, $\eta$ is the fraction of this magnetic energy in a field
directed along the rotation axis and $\xi$ is a dimensionless parameter
depending on the density stratification and the spatial distribution of the
perturbing forces in the convective layer of the star (it can be calculated from
the gyration radius and the apsidal motion constant using stellar models). Here
we shall use $\xi \simeq 5$. The rotational frequency of the outer layers of the
star is taken to be that of the orbit (caused by synchronisation) and in
Applegate's model ${\cal M} = 0$. Adopting a companion star radius $R =
0.064$\,R$_{\odot}$ yields: 
\[
Q = 3.4 \times 10^{47}\,{\rm g\,cm}^2.
\] 
Consequently, the quadrupole moment $Q$ of the star is one to two
orders of magnitude larger than the variations required for the
observed $P_{\rm b}$ changes. Using this value in
Equation~\ref{eq:soc} results in the angles $\theta$ required for the
observed $x$ variations. These values can in turn be used to correct
the $\Delta Q$ values shown in Table~\ref{tab:DQnot} for the angular
dependence, though this effect is all but negligible. The resulting
changes required in the tilt and quadrupole moment of the companion
star are collated in Table~\ref{tab:theta}.

\begin{table}
\center
\caption{Quadrupole changes (orientation and magnitude) that explain the
changes in $P_{\rm b}$ and $x$ for the five epochs.}
\begin{tabular}{rrr}
\hline {Epoch} &{$\theta$\,(deg)} & {$\Delta Q/Q$} \\ \noalign{\smallskip}
\hline 
\noalign{\smallskip} 
1 & $ -0.14$ & $+0.041 $ \\ 
2 & $ -0.64$ & $-0.017 $ \\ 
3 & $ +0.73$ & $-0.041 $ \\ 
4 & $ -0.06$ & $+0.052 $ \\ 
5 & $ +0.02$ & $-0.019 $ \\ 
\noalign{\smallskip} 
\hline
\end{tabular}
\label{tab:theta}
\end{table}

We conclude that with small changes in the orientation and magnitude
of the quadrupole moment, we can explain all observed parameter
changes. The change in the sign of $\dot x$ from one epoch to the
other can be easily explained by a small oscillation of the symmetry
axis of the quadrupole. Moreover, if we use an alternative scenario
with a larger filling factor for the radius of the companion star,
these numbers will get even smaller since $Q$ will be larger. We do
not, however, know of any physical mechanism that would cause the
required tilt of the star or its rapid changes.

\subsection{Alternative Models for the Gravitational Quadrupole-Moment
  Variations}\label{subsec:LanzaRodono}

The Applegate model is one of a number that have been applied to
explain the cyclic modulation in the orbital period of magnetically
active close binaries. In Applegate's model a rather large fraction of
the stellar luminosity is required for its operation. Hence, one might
be able to detect changes in the stellar luminosity in phase with the
orbital period modulation. The model of Lanza \& Rodon\'{o}
\citep[e.g.][]{lrr98,lr99b} considers variations in the azimuthal
B-field to explain the variations in oblateness. A change in the
azimuthal field intensity can also produce a change in the quadrupole
moment by changing the effective centrifugal acceleration. However,
one should keep in mind that it remains to be established what kind of
companion star we have in \psr.

\subsection{The Nature of the Companion Star}\label{subsec:companion}
There seem to be three possibilities for the nature of the companion
star:
\begin{enumerate}[I]
\item a white dwarf (WD), 
\item a brown-dwarf--like star and
\item a semi-degenerate helium star (He).
\end{enumerate}
In the following we briefly discuss each of these possibilities. 

For a WD companion star of $M_{\rm c} \geq 0.04\,M_{\odot}$, the
mass-radius relation of a non-relativistic degenerate Fermi-gas
\citep[e.g.][]{st83} yields an upper limit for the radius of $R_{\rm
  c} \leq 0.025\,$R$_{\odot}$. This value is much less than what is
estimated in optical since a combination of the minimum value for the
filling factor of 0.43 and the Roche-lobe size of $R_{\rm
  L}=0.15$\,R$_{\odot}$, yields a minimum optical radius of $R_{\rm c}
\geq 0.064$\,R$_{\odot}$. However, because of irradiation by the
pulsar wind and tidal dissipation of energy in the WD envelope it is
likely that the WD is bloated in size. The effect of a thermally
bloated WD in a close binary system has recently been detected in a
transiting source by the Kepler satellite \citep{crf11}. In this case
the young (few hundred Myr) hot white dwarf is bloated by a factor of
seven in size. The companion star in \psr\ is believed to be much
older, however it is quite possible that an extended H-rich atmosphere
may exist in this star from the effects mentioned above.  On the other
hand, the actual effect of the pulsar wind remains uncertain and may
not be as efficient as expected previously. For example, Equation~(2)
of \cite{tav92} for the irradiation of the companion as induced by the
pulsar wind, yields a surface temperature of the companion of
$6200$\,K. However, the observed temperature is $\leq 3000$\,K
\citep{svbk01b}.

Brown dwarf models have been applied \citep{bc01} to explain the
nature of the $\sim 0.05$\,M$_{\odot}$ donor star in the accreting
millisecond pulsar system SAX J1808.4-3658. Brown dwarfs have a
relation between mass and radius that is acceptable from observational
constraints, also for \psr. However, the obtained mass-radius relation
was based on brown-dwarf models by \cite{cbah00}, calculated for
isolated low-mass stars. We find it questionable if the present core
remnant in \psr\ can simply be described by the equivalent of a brown
dwarf. The original ZAMS progenitor system of \psr\ definitely did not
consist of a B-star and a brown dwarf: the initial sub-stellar mass of
a brown dwarf is $ < 0.08$\,M$_{\odot}$ (the limit for H-ignition) and
the progenitor of the neutron star had a mass $>
10$\,M$_{\odot}$. Hence, the initial binary would have had a mass
ratio of $< 1/100$, which is not only disfavoured by formation, but
also unlikely given that this binary should undergo a common envelope
phase followed by survival of a supernova explosion and finally spin
up the pulsar with so little mass. A more reasonable possibility is
that \psr\ (and many other eclipsing binary millisecond pulsars) is
the outcome of a short orbital period, low-mass X-ray binary with
un-evolved main sequence stars. Such systems have been studied in
detail by, e.g., \cite{ps88} and \cite{elc97}. The latter authors also
discussed the final fate of a system like \psr: an ultra-compact LMXB
where the companion fills its Roche lobe (again), because of orbital
angular momentum losses, and eventually may undergo tidal disruption
leaving behind a single or planetary millisecond pulsar \citep[e.g. a
system like PSR~B1257+12,][]{wf92}.

Finally, it should be noted that the companion star in \psr\ could be
closely related to the semi-degenerate helium star companions in
ultra-compact AM~CVn systems \citep[e.g.][]{it91,npvy01}. One example
is the system SDSS~J0926+3624 \citep{cml+10} which has a similar
companion mass of $0.035$\,M$_{\odot}$ ($P_{\rm b} =
28$\,min). Although the accreting compact object in AM~CVn systems is
a WD, a similar evolution is expected for systems with an accreting
neutron star \citep{sdv86}. Also, in this case the expected
mass-radius relations yield acceptable values for the radius in
\psr\ -- especially if one accepts the possibility that a
pulsar wind and tidal dissipation could lead to a somewhat bloated
radius.

\subsection{High-Energy Emission Prospects for
  \psr}\label{subsec:x-rays}
The detection of X-rays from \psr\ could yield vital information about
its orbital evolution, the state of the companion star and
interactions between the pulsar wind and the evaporated material of
the companion star. However, no X-rays are detected from \psr\ in the
RXTE All-Sky-Survey (A.M. Levine, private communication). The RXTE~ASM
sensitivity in the 2-10\,keV band is $\sim 30$\,mCrab, corresponding
to an energy flux of $F_{\rm x} \simeq 7 \times
10^{-10}$\,erg\,s$^{-1}$\,cm$^{-2}$.  Given the distance to \psr\ ($d
\simeq 1.04$\,kpc), we can estimate an upper limit to the mass
accretion rate:
\begin{equation}
  \label{Macc}
  \dot{m} \leq \frac{4\pi\,d^2\,F_{\rm x}\,R_{\rm p}}{G\,m_{\rm p}} 
\end{equation}
Assuming a neutron star mass $m_{\rm p}=1.8$\,M$_{\odot}$ and radius
$R_{\rm p}=10^6$\,cm, we obtain a conservative upper limit for the
potential accretion rate $\dot{m}\lesssim 6.0 \times
10^{-12}$\,M$_{\odot}$\,yr$^{-1}$. Keeping in mind that the expected
mass transfer rate for a Roche-lobe-filling $\sim 0.05$\,M$_{\odot}$
companion star is expected to be much lower than this limit -- and
given the fact that we do observe a radio pulsar -- it is not a
surprise that \psr\ is not detected in the RXTE ASM data. Chandra
and/or XMM data would impose far more stringent constraints; or might
even provide a detection.

One must bear in mind that the lack of detected X-rays does not
exclude mass loss from the companion star. It is possible that the
pulsar wind (which is probably enhanced in the equatorial region
towards the companion star) is able to prevent any accretion onto the
neutron star. However, in this scenario it needs to be investigated
thoroughly whether or not a shock front would lead to acceleration of
protons and subsequent production of high-energy $\gamma$-rays
\citep{hg90,sgk+03}. 

Considering $\gamma$-ray emission from this system, with a distance of
$\simeq1.04$\,kpc and a spin-down luminosity $\dot{E}=4\pi^2 I
\dot{P}/P^3=5.5\times 10^{33}$\,erg$/$s, this pulsar is a good
candidate for detection with the Large Area Telescope of the
\emph{Fermi} satellite \citep{aaa09}. However none of the \emph{Fermi}
sources in the first-year catalogue lie within a radius of $3^{\circ}$
of \psr\ \citep{aaa10}. It would therefore be interesting to see if
\psr\ is detected by \emph{Fermi} with accumulated data.

\section{Summary}\label{sec:conclusion}
We have presented a timing update on \psr\ and shown that the
variations of the orbital period and projected semi-major axis of the
binary system are far more extreme than described in earlier work. We
have analysed all possible causes for these variations and found that
a combination of gravitational quadrupole coupling and spin-orbit
coupling are the most likely origins of this behaviour, though
accurate modelling efforts are still required. We furthermore
discussed the nature of the companion star and conclude that a
semi-degenerate helium star is more likely than the white-dwarf
alternative and, finally, we expect high-energy $\gamma$-rays may be
visible in this system if a pulsar wind prevents accretion onto the
neutron star -- which in turn is expected based on the lack of
X-rays. 

\section*{Acknowledgements}
We are very grateful to all staff at the Effelsberg, Westerbork,
Jodrell Bank and Nan\c cay radio telescopes for their help with the
observations used in this work. Part of this work is based on
observations with the 100-m telescope of the Max-Planck-Institut f\"ur
Radioastronomie (MPIfR) at Effelsberg. Access to the Lovell telescope
is supported through an STFC rolling grant. The Nan\c cay radio
telescope is part of the Paris Observatory, associated with the Centre
National de la Recherche Scientifique (CNRS) and partially supported
by the R\'egion Centre in France. The Westerbork Synthesis Radio
Telescope is operated by the Netherlands Foundation for Research in
Astronomy (ASTRON) with support from the NWO. The Parkes radio
telescope is part of the Australia Telescope National Facility which
is funded by the Commonwealth of Australia for operation as a national
facility managed by CSIRO. JPWV is supported by the European Union
under Marie-Curie Intra-European Fellowship 236394.  We want to thank
Alan M. Levine (MIT) and Diego Altamirano (University of Amsterdam)
for their help with investigating RXTE data, Lucas Guillemot (MPIfR)
for discussions on Fermi results and Paulo Freire (MPIfR) for
discussions on black widow and eclipsing systems. Finally, we thank
the referee, Ingrid Stairs (UBC), for helpful and interesting
comments.

\bibliographystyle{mn2e}
\bibliography{journals,psrrefs,modrefs,crossrefs}

\end{document}